\documentclass[twocolumn]{autart}

\usepackage{graphicx}
\usepackage{caption}
\usepackage[T1]{fontenc}
\usepackage{amsmath,amssymb}
\usepackage{amsfonts}

\newcommand{\E}[2]{\mathbb{E}_{#1}\left[#2\right]}

\newtheorem{Theorem}{Theorem}[section]

\newtheorem{Lemma}[Theorem]{Lemma}
\newtheorem{Definition}[Theorem]{Definition}
\newtheorem{Remark}[Theorem]{Remark}
\newtheorem{Example}[Theorem]{Example}
\newtheorem{Assumption}[Theorem]{Assumption}

\usepackage[usenames,dvipsnames]{xcolor}
\definecolor{wheat}{rgb}{0.96,0.87,0.70}
\definecolor{mario}{rgb}{0.8,0.8,1}
\definecolor{seb}{rgb}{0.8,1,0.8}
\definecolor{myGreen}{rgb}{0.,0.8,0.0}

\definecolor{darkgreen}{rgb}{0,0.6,0}

\newcommand {\matr}[2]{\left[\begin{array}{#1}#2\end{array}\right]}

\newcounter{lastnote}

\begin{document} 

\begin{frontmatter}

\title{Economic Linear Quadratic MPC With Non-Unique Optimal Solutions\thanksref{footnoteinfo}}
\thanks[footnoteinfo]{This paper was not presented at any IFAC 
	meeting. Corresponding author M. Zanon.}
	
\author[Mario]{Mario Zanon}\ead{mario.zanon@imtlucca.it}
\address[Mario]{IMT School for Advanced Studies Lucca, Piazza San Francesco 19, 55100, Lucca, Italy}

\begin{abstract}
	Asymptotic stability in economic receding horizon control can be obtained under a strict dissipativity assumption, related to positive-definiteness of a so-called rotated cost, and through the use of suitable terminal cost and constraints. In the linear-quadratic case a common assumption is that the rotated cost is positive definite. The positive semi-definite case has received surprisingly little attention, and the connection to the standard dissipativity assumption has not been investigated. In this paper, we fill this gap by connecting existing results in economic model predictive control with the stability results for the semi-definite case, the properties of the constrained generalized discrete algebraic Riccati equation, and of two optimal control problems. Moreover, we extend recent results relating exponential stability to the choice of terminal cost in the absence of terminal constraints.
\end{abstract}

\begin{keyword}
	Economic model predictive control; dissipativity; exponential stability; constrained generalized Riccati equations.
\end{keyword}

\end{frontmatter}

\section{Introduction}

Model Predictive Control (MPC) is a control technique in which a finite-horizon Optimal Control Problem (OCP) is solved at each time step. The first optimal control input is applied, the next state is measured, and the OCP is solved again in order to update the optimal control input in a receding horizon fashion~\cite{Rawlings2017,Gruene2017}.

Though the stability theory typically assumes that the cost function has its minimum at the optimal steady state, in economic MPC (EMPC) generic costs are allowed. While this allows one to potentially improve closed-loop performance, providing asymptotic stability guarantees becomes much harder and typically requires some strict dissipativity assumption to hold~\cite{Faulwasser2018a}. Early results include~\cite{Diehl2011,Amrit2011a}, where the main idea was first presented. Necessity of dissipativity for optimal steady state operation has been proven in~\cite{Mueller2015,Mueller2015a}. Stability in the absence of terminal constraints has been analyzed in~\cite{Grune2013a,Zanon2018a,Faulwasser2018,Zanon2025,Gruene2025}. The periodic case has been analyzed in~\cite{Zanon2017e,Mueller2016} and the discounted case has been studied in~\cite{Gaitsgory2018,Zanon2022a}. 
Computationally efficient algorithms tailored to EMPC have been proposed in~\cite{Quirynen2016,Quirynen2017a,Verschueren2017,Zanon2021b}. 

The linear-quadratic case has been analyzed in~\cite{Zanon2014d,Zanon2016b,Zanon2017a,Facchino2023}, where the use of a quadratic positive-definite cost has also been proposed to formulate tracking MPC so as to approximate EMPC. While a positive-definite rotated cost is assumed in these works, the relaxation of that assumption has received less attention. The only work we are aware of, which only assumes that the rotated stage cost is positive semi-definite is~\cite{Olanrewaju2017}, which has received surprisingly little attention. In~\cite{Olanrewaju2017}, exponential stability is proven by relying on a seemingly non-strict dissipativity assumption. However, while the standard strict dissipativity assumption is known to yield the same result, the connection between these two (at first sight) rather different assumptions has never been investigated before, to the best of the author's knowledge. As we will prove in this paper, the two assumptions actually imply each other. Moreover, the connection between them can be done through the formulation of two infinite-horizon OCPs, one causal and forward in time, while the other one anticausal and backward in time. This result further highlights the deep connection between dissipativity and OCPs formulated in forward and backward time, as discussed in the seminal papers on dissipativity~\cite{Willems1971,Willems1972a,Willems1972b} (though the concept of strict dissipativity has been introduced only much later). Finally,~\cite{Zanon2025} investigates how exponential stability can be obtained by means of a suitable terminal cost in finite horizon linear quadratic EMPC under the requirement that the rotated cost is positive definite. We will extend those results to the positive semi-definite case. 

Our main contributions are as follows. We obtain a new asymptotic stability proof for linear-quadratic MPC under quadratic strict pre-dissipativity; moreover, we show that this assumption is fully equivalent to both the one used in~\cite{Olanrewaju2017} in a similar context and the standard strict dissipativity assumption used in economic MPC. Additionally, we extend the stability results to the case of a finite horizon and provide explicit bounds on the terminal cost. Finally, we briefly discuss the case in which some but not all optimal solutions are stabilizing.

The main difficulty when relaxing the requirement on the cost to be positive definite, which entails that the solution is unique, is that the OCP solution is not related to the Discrete Algebraic Riccati Equation (DARE). Nevertheless, it can be related to a Constrained Generalized DARE (CGDARE), see~\cite{Ferrante2013} which provides an excellent discussion and a very sharp analysis of the properties of the CGDARE and its solutions, their relation to the corresponding infinite-horizon optimal control problem and to exponential stability. We will exploit those results and extend them to prove our claims.

The remainder of this paper is structured as follows. In Section~\ref{sec:problem_statement} we state the optimal control problem and the corresponding Riccati equations. In Section~\ref{sec:preliminary} we provide some preliminary results that will be helpful to prove our main contribution which is stated in Section~\ref{sec:results} for the infinite-horizon case and in Section~\ref{sec:terminal_cost} for the finite horizon case. We briefly comment on the case in which some optimal controls might be stabilizing while others are not in Section~\ref{sec:difficult_case} and we conclude the paper in Section~\ref{sec:conclusions}.

\section{Problem Statement}
\label{sec:problem_statement}

Consider the linear system
\begin{align*}
	x_{k+1} &= A x_k  + B u_k,
\end{align*}
with state $x\in\mathbb{R}^{n_x}$ and control input $u\in\mathbb{R}^{n_u}$; 
and the following linear-quadratic Optimal Control Problem (OCP)
\begin{subequations}
	\label{eq:ocp}
	\begin{align}
		V(\hat x_0) = \min_{x,u} \ & \lim_{N\to\infty}\sum_{k=0}^{N-1} \matr{c}{x_k \\ u_k}^\top H \matr{c}{x_k \\ u_k} + x_N^\top P^\mathrm{f} x_N \\
		\mathrm{s.t.} \ & x_0 = \hat x_0, \\
		&x_{k+1} = A x_k + B u_k,
	\end{align}
\end{subequations}
with symmetric $H$ and $P^{\mathrm{f}}$, and
\begin{align*}
	H := \matr{ll}{Q & S^\top \\ S & R}.
\end{align*}

Under suitable assumptions on $P^\mathrm{f}$, the LQR solution is characterized by a solution $P$ with associated feedback matrix $K$ of the Constrained Generalized Discrete Algebraic Riccati Equation (CGDARE)~\cite{Ferrante2013}
\begin{subequations}
	\label{eq:CGDARE}	
	\begin{align}
		P =\ & Q + A^\top P A - (S^\top + A^\top P B)K, \\
		K =\ & (R+B^\top P B)^{\dagger} (S+B^\top P A), \\
		&\mathrm{ker}(R + B^\top P B) \subseteq \mathrm{ker}(S^\top + A^\top P B), \label{eq:CGDARE:kernel}
	\end{align}
\end{subequations}
where $\dagger$ denotes the Moore-Penrose pseudo-inverse. Note that, as discussed in~\cite{Ferrante2013}, Condition~\eqref{eq:CGDARE:kernel} is necessary in order for the CGDARE to correctly relate to OCP~\eqref{eq:ocp}, as it is directly related to the optimality conditions of OCP~\eqref{eq:ocp}. Furthermore, we are interested in symmetric solutions, and we will simply refer to symmetric solutions as solutions in the remainder of this paper.

For any (symmetric) matrix $M$, we define
\begin{subequations}
	\label{eq:cost_rotation}
	\begin{align}
		Q_M &:= Q + A^\top M A - M, & S_M &:= S + B^\top M A, \\ R_M &:= R + B^\top M B, & H_{M}&:=\matr{ll}{Q_{M} & S_{M}^\top \\ S_{M} & R_{M}},
	\end{align}
\end{subequations}
such that the CGDARE can be written as
\begin{subequations}
	\label{eq:CGDARE_compact}	
	\begin{align}
		0 =\ & Q_P- S_P^\top K, \\
		K =\ & R_P^{\dagger} S_P, \\
		\mathrm{ker}(R_P) \subseteq \ & \mathrm{ker}(S_P^\top).
	\end{align}
\end{subequations}
We will also use the same shorthand notation in order to define various dissipativity assumptions.

As proven in~\cite{Ferrante2013}, the CGDARE provides the optimal feedback law
\begin{align}
	\label{eq:optimal_feedback}
	F(x,v) = -Kx + Gv, && v \text{ arbitrary},
\end{align}
where
\begin{align*}
	G = I-R_P^\dagger R_P,
\end{align*}
defining the closed-loop system
\begin{align}
	\label{eq:optimal_closed_loop}
	x_+ = (A-BK)x + BGv, && v \text{ arbitrary},
\end{align}
and the quadratic value function $V(x) = x^\top P x.$

In this paper, we are interested in studying under which conditions Problem~\eqref{eq:ocp} yields a feedback law~\eqref{eq:optimal_feedback} which renders the closed-loop system~\eqref{eq:optimal_closed_loop} asymptotically stable. General results have been obtained in the literature~\cite{Faulwasser2018a,Diehl2011,Amrit2011a,Mueller2015,Grune2013a,Zanon2018a,Faulwasser2018,Zanon2025,Gruene2025,Zanon2017e,Mueller2016,Gaitsgory2018,Zanon2022a,Zanon2014d,Zanon2016b,Zanon2017a,Facchino2023} by relying on the concept of strict dissipativity, where the main idea is that a positive-definite cost (called the rotated cost) allows one to prove the existence of a hidden tracking structure, hence obtaining a Lyapunov function for the economic MPC controller. 
For the linear-quadratic case, the rotated stage cost is characterized, by matrix $H_\Lambda$, for a given matrix $\Lambda$ as per~\eqref{eq:cost_rotation}. As observed in~\cite{Zanon2016b}, strict dissipativity can be checked by solving a Semidefinite Program (SDP) in case the solution is unique. In this paper, we aim at extending that result to the case in which the MPC solution is not unique, i.e., in case $G\neq0$. As it will become clear throughout the paper, that endeavor is not trivial and requires some care.

We introduce next a few useful definitions and results that we will exploit in order to derive our main results. In case $R_P$ is invertible, then the CGDARE coincides with the Discrete Algebraic Riccati Equation (DARE)~\cite{Ferrante2013}
\begin{subequations}
	\label{eq:DARE_compact}	
	\begin{align}
		0 =\ & Q_P- S_P^\top K, \\
		K =\ & R_P^{-1} S_P,
	\end{align}
\end{subequations}
where, since $R_P$ is full rank $\mathrm{ker}(R_P)=0$, and the last condition in~\eqref{eq:CGDARE_compact} is automatically satisfied. Throughout the paper, we will also exploit the properties of the regularized DARE (rDARE)
\begin{subequations}
	\label{eq:rDARE_compact}	
	\begin{align}
		0 =\ & Q_P- S_P^\top K, \\
		K =\ & (R_P+G)^{-1} S_P.
	\end{align}
\end{subequations}
If it exists, we will denote the unique stabilizing solution of the CGDARE or rDARE as $P_\mathrm{s}$, with corresponding feedback $K_\mathrm{s}$ such that the eigenvalues of $A-B K_\mathrm{s}$ are all inside the unit circle. We will prove later that, under conditions that will turn out to be necessary for stability, the solutions of the CGDARE and rDARE coincide such that they do not need to be distinguished.

In order to prove our results, we will also exploit the Reverse Constrained Generalized Discrete-time Algebraic Riccati Equation (RCGDARE), defined as:
\begin{subequations}
	\begin{align}
		\bar P &= \bar Q + \bar A^\top \bar P \bar A - (\bar S^\top + \bar A^\top \bar P \bar B) \bar K, \\
		\bar K &= (\bar R + \bar B^\top \bar P \bar B)^{\dagger}(\bar S + \bar B^\top \bar P \bar A),\\
		&\mathrm{ker}(\bar R+\bar B^\top \bar P \bar B) \subseteq \mathrm{ker}(\bar S+\bar B^\top \bar P \bar A),
	\end{align}
\end{subequations}
where 
\begin{align*}
	\bar A &:= A^{-1}, & \bar B &:= A^{-1} B, \\
	\bar Q &:= -\bar A^\top Q \bar A, & \bar S &:= S \bar A - \bar B^\top Q \bar A, \\
	\bar R &:= -R + S \bar B + \bar B^\top S^\top - \bar B^\top Q \bar B, \hspace{-10em} \\
	\bar H &:= \matr{ll}{\bar Q & \bar S^\top \\ \bar S & \bar R}.
\end{align*}
If it exists, we define as $\bar P_\mathrm{s}$ the stabilizing solution of the RCGDARE (and the corresponding rRDARE). If $R-SA^{-1}B$ is nonsingular, this solution coincides with the antistabilizing solution of the CGDARE/rDARE~\cite{Ionescu1996}. This establishes a strong similarity with the continuous-time case, where the antistabilizing solution is the stabilizing solution of the problem formulated in reverse time.

\begin{Remark}
	Though throughout the paper we assume that $(A,B)$ is stabilizable, we will first restrict our attention to the study of its controllable subpart. In that case, we will exploit the fact that for any controllable system, one can introduce feedback matrix $F$ which makes $A-BF$ full rank. Consequently, as discussed in, e.g.,~\cite[Theorem~3.8, Remark~3.9]{Zanon2025}, the existence of $A^{-1}$ can be assumed without loss of generality in the controllable case. After deriving the results for the controllable case, we will generalize them to the stabilizable case.
\end{Remark}

\section{Preliminary Results}
\label{sec:preliminary}

Before proving our main results, we state the following useful properties of the CGDARE.

We define $Z$ as the null space matrix of $R_P$, i.e.
\begin{align}
	\label{eq:Z_definition}
	R_PZ&=0, & Z^\top Z&=I.
\end{align}

\begin{Lemma}
	\label{lem:G_Z}
	Matrix $G$ satisfies
	\begin{align*}
		G=ZZ^\top,
	\end{align*}
	where $Z$ is defined as per~\eqref{eq:Z_definition}.
\end{Lemma}
\begin{pf}
	We observe that a singular value decomposition on the symmetric matrices $R_P$ and $R_P^\dagger$ yields
	\begin{align}
		G &= I - R_P^\dagger R_P \nonumber \\
		&= I - V \Sigma^\dagger V^\top V \Sigma V^\top \nonumber \\
		&= V V^\top - V \Sigma^\dagger  \Sigma V^\top. \label{eq:G_svd}
	\end{align}
	Note that $V =\matr{cc}{Y & Z}$ is orthonormal, where $Y$ and $Z$ are, respectively, range and null space matrices of $R_P$ and, consequently, also of $R_P^\dagger$. Moreover, 
	\begin{align*}
		\Sigma^\dagger  \Sigma &= \matr{rr}{I & 0 \\ 0 & 0}, & I-\Sigma^\dagger  \Sigma &= \matr{rr}{0 & 0 \\ 0 & I}.
	\end{align*}
	Consequently, we have that~\eqref{eq:G_svd} reads
	\begin{align*}
		G = ZZ^\top,
	\end{align*}
	where, in case $R_P$ is full rank,  $\Sigma^\dagger  \Sigma =I$, $Z=0$, and $G=0.$
	$\hfill\qed$
\end{pf}
We recall that, for $P_1,P_2$ solutions of the CGDARE,~\cite[Theorem~4.3]{Ferrante2013} states that $\ker R_{P_1}=\ker R_{P_2}$. Consequently, matrix $G$ is independent of which $P$ solving the CGDARE is used to define it. 

\begin{Lemma}
	\label{lem:G_dare}
	For any $P$ solving the CGDARE it holds that
	\begin{align*}
		(R_P+G)^{-1} &= R_P^\dagger + G, \\
		(R_P+G)^{-1} S_P &= R_P^\dagger S_P.
	\end{align*}
	Consequently, any solution $P$ with corresponding feedback $K$ of the CGDARE is a solution of the rDARE, though the converse might not be true.
\end{Lemma}
\begin{pf}
	We observe that
	\begin{align*}
		R_P+G &= R_P+I-R_P^\dagger R_P \\
		&= V\Sigma V^\top + VV^\top - V\Sigma^\dagger \Sigma V^\top \\
		&= V \left (\matr{cc}{\Sigma_+ & 0 \\ 0 & 0} + \matr{cc}{0 & 0 \\ 0 & I}\right ) V^\top,
	\end{align*}
	with $\Sigma_+\succ0$, such that
	\begin{align*}
		(R_P+G)^{-1} &= V \left (\matr{cc}{\Sigma_+^{-1} & 0 \\ 0 & 0} + \matr{cc}{0 & 0 \\ 0 & I}\right ) V^\top \\
		&= R_P^\dagger + G.
	\end{align*}
	Moreover, since $S_P^\top G=0$, we have that $GS_P=0$, which proves that $P$, $K$ are a solution of the rDARE.
	
	Finally, we observe that, for all $P$ solving the CGDARE one can simplify the rDARE as
	\begin{align*}
		0 &= Q_P- S_P^\top K, &
		K &= (R_P+G)^{-1} S_P\\
		&&&=R_P^\dagger S_P.
	\end{align*}
	However, the rDARE can have additional solutions $P_\mathrm{rDARE}$ which do not satisfy 
	\begin{align*}
		R_{P_\mathrm{rDARE}}G&=0, & 	S_{P_\mathrm{rDARE}}^\top G&=0,
	\end{align*}
	which can then not be solutions of the CGDARE.
	$\hfill\qed$
\end{pf}
In order to further clarify the result of Lemma~\ref{lem:G_dare}, and its implications on the connection between solutions of the CGDARE and the rDARE, let us introduce the following example.

\begin{Example}
	Consider a system with matrices
	\begin{align*}
		A&=\matr{rr}{2&1\\0&1}, & B&=\matr{rr}{2&0\\1&1},
	\end{align*}
	with stage cost matrices
	\begin{align*}
		Q&=\matr{rr}{0&0\\0&1}, & R&=0, &S&=0.
	\end{align*}
	The CGDARE has solution
	\begin{align*}
		P&=\matr{rr}{0&0\\0&1}, & K&=\frac{1}{2}\matr{rr}{0&1\\0&1}, &
		G&=\frac{1}{2}\matr{rr}{1&-1\\-1&1}.
	\end{align*}
	The rDARE has a stabilizing solution
	\begin{align*}
		P_\mathrm{s}&=I, & K_\mathrm{s}&=\matr{rr}{0.75&0.5\\-0.25&0.5}.
	\end{align*}
	One can verify that $P$ solves the rDARE, though it is not the stabilizing solution. Finally, one can also check that $P_\mathrm{s}$ does not solve the CGDARE. Note that for this example we have
	\begin{align*}
		BG = \matr{rr}{1&-1\\0&0},
	\end{align*}
	i.e., the closed-loop behavior is not unique, as different optimal control inputs yield different state trajectories.
\end{Example}

This example motivates a deeper study of the case $BG=0$. We will prove in Section~\ref{sec:results} that, under a quadratic strict pre-dissipativity assumption we can guarantee that $BG=0$. In the next lemma, we prove that $BG=0$ entails that the solutions of the CGDARE coincide with the solutions of the rDARE. 

\begin{Lemma}
	\label{lem:BG0_rdare}
	Assume that $BG=0$ and $R_P\succeq0$. Then any $P$ solving the CGDARE is a solution of the rDARE and vice versa.
\end{Lemma}
\begin{pf}
	We observe that $BG=0$ entails $RG=0$ and $S^\top G=0$, such that for all symmetric $\Lambda$ it holds that, from Equation~\eqref{eq:cost_rotation}, we have
	\begin{align*}
		R_\Lambda G&= 0, & S_\Lambda^\top G&=0.
	\end{align*}
	Consequently, for any $\Lambda$ such that $R_\Lambda\succeq 0$ we have
	\begin{align}
		(R_\Lambda+G)^{-1} &= R_\Lambda^\dagger + G, &
		(R_\Lambda+G)^{-1}S_\Lambda &= R_\Lambda^\dagger S_\Lambda.
		\label{eq:rDARE_CGDARE}
	\end{align}
	Because $R_P\succeq0$ for all $P$ solving the CGDARE, then Lemma~\ref{lem:G_dare} entails that $R_P+G\succ0$ solves the rDARE for all $P$ solving the CGDARE. Consequently, $R_P+G\succ0$ for all $P$ solving the CGDARE as they also solve the rDARE, then $R_P+G\succ0$ for all $P$ solving the rDARE. In turn, this entails that $R_P\succeq0$ for all $P$ solving the rDARE. Consequently,~\eqref{eq:rDARE_CGDARE} holds for all $P$ solving the rDARE, which entails that all $P$ solving the rDARE also solve the CGDARE. $\hfill\qed$
\end{pf}

\begin{Lemma}
	\label{lem:BG0}
	If any $P$ solving the CGDARE with $H\succeq0$ is full rank, then $BG=0$.
\end{Lemma}
\begin{pf}
	By~\cite[Theorem~4.3]{Ferrante2013}, $G$ is independent of which symmetric $P$ solving the CGDARE is used to define it.
	Moreover, by~\cite[Lemmas~4.1,~4.2]{Ferrante2013}, in case $H\succeq0$
	\begin{align*}
		\ker R_P = \ker \matr{c}{PB \\ R},
	\end{align*}
	for any $P$ solving the CGDARE.
	Because we assume that $P$ is full rank, we further have
	\begin{align}
		\label{eq:ker_RB}
		\ker R_P = \ker \matr{c}{PB \\ R}= \ker \matr{c}{B \\ R}. 
	\end{align}
	
	Lemma~\ref{lem:G_Z} yields $G=ZZ^\top$. 
	Moreover, by~\eqref{eq:ker_RB}, $Z$ is a null space matrix for $B$ and  $BG=BZZ^\top = 0.$ $\hfill\qed$
\end{pf}

Because it is easier to study the properties of OCP~\eqref{eq:ocp} and the corresponding CGDARE in case $H\succeq 0$, we will exploit so-called cost rotations to reconduct the generic case to the case $H\succeq0$. For any matrix $\Lambda$, these cost rotations consist in replacing the stage and terminal cost matrices $H$ and $P^\mathrm{f}$ with $H_\Lambda$ defined as per~\eqref{eq:cost_rotation} and $P^\mathrm{f}_\Lambda := P^\mathrm{f} - \Lambda$. 

\begin{Remark}
	In the literature on economic MPC, cost rotations for a given matrix $\Lambda$ are usually defined as $H_{-\Lambda}$ and $P^\mathrm{f}_\Lambda := P^\mathrm{f} + \Lambda$. Since that is just a convention and in the context of this paper it is notationally and conceptually simpler to use $H_{\Lambda}$ and $P^\mathrm{f}_\Lambda := P^\mathrm{f} - \Lambda$, we will stick to this last convention.
\end{Remark}

We will prove in Section~\ref{sec:results} that assuming the existence of matrix $\Lambda$ such that $H_\Lambda\succeq0$ is not restrictive. We provide next some useful properties related to cost rotations.

We define the rotated linear-quadratic OCP as
\begin{subequations}
	\label{eq:ocp_rotated}
	\begin{align}
		V_\Lambda(\hat x_0) = \min_{x,u} \ & \lim_{N\to\infty}\sum_{k=0}^{N-1} \matr{c}{x_k \\ u_k}^\top H_\Lambda \matr{c}{x_k \\ u_k} + x_N^\top P^\mathrm{f}_\Lambda x_N \\
		\mathrm{s.t.} \ & x_0 = \hat x_0, \\
		&x_{k+1} = A x_k + B u_k,
	\end{align}
\end{subequations}
which only differs from the original OCP~\eqref{eq:ocp} in that $H$ and $P^\mathrm{f}$ are replaced by $H_\Lambda$ and $P^\mathrm{f}_\Lambda=P^\mathrm{f}-\Lambda$.

\begin{Lemma}
	\label{lem:cost_rotation_equivalence}
	The costs of the rotated OCP~\eqref{eq:ocp_rotated} and the original OCP~\eqref{eq:ocp} only differ by the constant term $x_0^\top \Lambda x_0$ and the cost-to-go matrix of the rotated OCP satisfies $P_\Lambda=P-\Lambda$. Moreover, any state-input trajectory that is optimal for the rotated OCP is also optimal for the original OCP and vice versa.
\end{Lemma}
\begin{pf}
	The first claim is proven in~\cite{Amrit2011a,Zanon2025} or \cite[Lemma~2.2]{Olanrewaju2017}, where it is observed that for feasible trajectories the costs only differ by the constant term $\hat x_0^\top \Lambda \hat x_0$. The second claim directly follows. $\hfill\qed$
\end{pf}

We recalled before that $G$ is independent of which CGDARE solution $P$ is used to define it. We prove next that it is also independent of cost rotations.

\begin{Lemma}
	\label{lem:G_indep_rotations}
	Matrix $G$ is independent of cost rotations.
\end{Lemma}
\begin{pf}
	We exploit the fact that $P_\Lambda=P-\Lambda$ to write
	\begin{align*}
		R_\Lambda + B^\top P_\Lambda B &= R + B^\top \Lambda B + B^\top (P-\Lambda) B \\
		&= R + B^\top P B. 
	\end{align*}
	Consequently, $G$ is independent of cost rotations. $\hfill\qed$
\end{pf}

We prove next a result related to the RCGDARE which will be useful later on.
\begin{Lemma}
	\label{lem:barG=G}
	Suppose that $BG=0$ 
	and $A^{-1}$ exists, such that the RCGDARE is well defined. Then, $\bar G = G$.
\end{Lemma}
\begin{pf}
		Because $BG=0$ we have $RG = 0$, $S^\top G = 0$.
		Consequently, we immediately obtain 
		\begin{align}
			\bar BG &= 0, & \bar RG &= 0, & \bar S^\top G &= 0.
		\end{align}
		Since $\bar B\bar G = 0$ entails $\bar R\bar G = 0$,  $\bar S^\top \bar G = 0$, we immediately obtain 
		\begin{align}
			B\bar G &= 0, & R\bar G &= 0, & S^\top \bar G &= 0.
		\end{align}
		Consequently, $\bar G = G$ and $\ker \bar R_{\bar P} = \ker R_P$.
	$\hfill\qed$
\end{pf}

Finally, as it will be helpful to discuss how our results are positioned in the more general context of nonlinear systems with generic stage costs, we recall next the common assumption used in the generic setting. For a given stage cost $\ell(x,u)$ and system dynamics $x_+=f(x,u)$, strict dissipativity with respect to the optimal steady state $x_\mathrm{s},u_\mathrm{s}$, i.e., the one optimizing $\ell$, is defined as follows. Note that in this paper we implicitly assume without loss of generality that $x_\mathrm{s}=0$, $u_\mathrm{s}=0$, $\ell(x_\mathrm{s},u_\mathrm{s})=0$.
\begin{Definition}[Strict Dissipativity]
	\label{def:nl_strict_diss}
	There exists a function $\lambda$ such that
	\begin{align}
		\label{eq:nl_strict_dissipativity}
		\mathcal{L}(x,u) &:= \ell(x,u) - \ell(x_\mathrm{s},u_\mathrm{s}) + \lambda(x) - \lambda(f(x,u)) \nonumber\\
		&\geq \rho(\|x-x_\mathrm{s}\|),
	\end{align}
	for all $x,u\in\mathcal{H}\subset \mathbb{R}^{n_x+n_u}$, with $\mathcal{H}$ compact and $\rho$ a positive definite function.
\end{Definition}
Note that $\mathcal{L}$ is referred to as rotated cost and $\rho$ is positive definite with respect to $x$, but not $u$, such that in the linear-quadratic case we call such a function positive semi-definite. The weaker dissipativity assumption is obtained if~\eqref{eq:nl_strict_dissipativity} holds with $\rho(\cdot)=0$. It has been proven in~\cite{Amrit2011a} that strict dissipativity is sufficient for asymptotic stability, where the rotated cost was used to prove that the economic MPC problem has a hidden tracking nature, which becomes explicit when the rotated cost $\mathcal{L}$ is used. Necessity of strict dissipativity for optimal steady state operation has been proven in~\cite{Mueller2015,Mueller2015a} under a local controllability assumption. A complete discussion on this subject can be found in, e.g.,~\cite{Faulwasser2018a}.

\section{Main Results}
\label{sec:results}

In this section, we first connect the existence of a solution to the CGDARE with dissipativity concepts. Subsequently, we connect the existence of a stabilizing solution to the CGDARE with strict dissipativity concepts. While the first part is mostly known, we find it useful to present the existing results with a point of view which better connects them with the second part of the paper and further clarifies the relationship with concepts commonly used in the literature about EMPC. The second part recalls some existing results, connects them with existing results stated for the general nonlinear case and further explains the deep connection of strict dissipativity and Riccati equations. 

\subsection{Existence of the Solution}
In order to discuss the existence of a solution to the CGDARE and the corresponding OCP~\eqref{eq:ocp}, we introduce the following dissipativity assumption.
\begin{Assumption}[Quadratic Pre-Dissipativity]
	\label{ass:dissipativity}
	There exists a symmetric matrix $\Lambda$ such that
	\begin{align*}
		H_{\Lambda} &\succeq 0.
	\end{align*}
\end{Assumption}

While in the next theorem we assume stabilizability, that could be relaxed by assuming that the uncontrollable unstable modes yield zero cost. However, as the focus of our analysis is proving exponential stability, our stronger assumption is not restrictive in the context of this paper.

\begin{Theorem}
	\label{thm:existence}
	Suppose that Assumption~\ref{ass:dissipativity} holds and $(A,B)$ is stabilizable. Then the CGDARE formulated with the rotated stage cost matrix $H_\Lambda\succeq0$ has a solution $P_\Lambda\succeq0$, which is also the cost-to-go of the infinite-horizon rotated OCP with terminal cost $P^\mathrm{f}_\Lambda = 0$ whose optimal feedback law is given by
	\begin{align*}
		u &= - Kx + Gv, & v \ \text{arbitrary}.
	\end{align*}
	The solution of the problem formulated using the original cost yields the same state and control input trajectories as the one of the rotated problem. The cost-to-go matrix of the original OCP satisfies
	\begin{align*}
		P = P_\Lambda + \Lambda.
	\end{align*}
\end{Theorem}
\begin{pf}
	The result is a consequence of~\cite[Theorem~2.1]{Ferrante2013}. The key observation is that the rotated problem~\eqref{eq:ocp_rotated} and the original problem~\eqref{eq:ocp} deliver the same feedback law and a shifted cost-to-go matrix $P_\Lambda=P-\Lambda$. Then, as proven in~\cite{Rappaport1971,Ferrante2013}, $P^\mathrm{f}_\Lambda=0$ guarantees that the optimal solution is obtained from the generalized Riccati difference equation (i.e., by value iteration). 
	One can then apply~\cite[Theorem~2.1]{Ferrante2013} to the rotated cost, since stabilizability of $(A,B)$ entails the existence of an infinite sequence of inputs yielding a finite rotated cost. 
	
	The optimal state and input trajectories of the original and rotated OCP coincide as proven in Lemma~\ref{lem:cost_rotation_equivalence}.
	Finally, one can write the rotated CGDARE as
	\begin{align*}
		0 =\ & Q_{P_\Lambda+\Lambda}- S_{P_\Lambda+\Lambda}^\top K, \\
		K =\ & R_{P_\Lambda+\Lambda}^{\dagger} S_{P_\Lambda+\Lambda}, \\
		\mathrm{ker}(R_{P_\Lambda+\Lambda}) \subseteq \ & \mathrm{ker}(S_{P_\Lambda+\Lambda}^\top).
	\end{align*}
	By replacing $P=P_\Lambda+\Lambda$ one immediately recovers the original CGDARE.
	$\hfill\qed$
\end{pf}
Note that the (in general non-unique) solutions of the CGDARE are independent of the terminal cost, whose proper choice is a necessary condition for the positive semi-definite solution to be the optimal cost, but not for its existence. Additionally, if the terminal cost matrix $P^\mathrm{f}$ is selected as any solution of the CGDARE, then the optimal cost-to-go matrix matches that specific solution of the CGDARE. A detailed analysis of the role of the terminal cost can be found in~\cite{Zanon2025} for the case in which quadratic strict $(x,u)$ pre-dissipativity holds, i.e., if there exists $\Lambda$ such that $H_\Lambda\succ0$. The extension of these results to the weaker quadratic strict pre-dissipativity Assumption~\ref{ass:stability_pd}-\ref{ass:strict_dissipativity} will be discussed in Section~\ref{sec:terminal_cost}.

Before moving forward with our analysis, we would like to discuss the necessity of Assumption~\ref{ass:dissipativity}. To that end, we recall that, as discussed in~\cite{Stoorvogel1998}, if any solution $P$ of the CGDARE satisfies $R_P\succeq0$, then all solutions satisfy the same inequality. Note that Theorem~\ref{thm:existence} establishes that both $R_\Lambda\succeq0$ and $P_\Lambda\succeq0$, such that also $R_P=R_{P_\Lambda}\succeq0$.

\begin{Theorem}
	\label{thm:dissipativity_necessity}
	Assume that the CGDARE has a symmetric solution $P$ satisfying $R_P\succeq0$. Then Assumption~\ref{ass:dissipativity} holds.
\end{Theorem}
\begin{pf}
	Since by assumption $R_P\succeq0$ and $\mathrm{ker}(R_P) \subseteq \mathrm{ker}(S_P^\top)$, then
	\begin{align*}
		H_P&\succeq0 && \Leftrightarrow & Q_{P} -S_{P}^\top R_{P}^{\dagger} S_{P} &\succeq 0.
	\end{align*}
	Since $P$ solves the CGDARE, we have
	\begin{align*}
	Q_{P} -S_{P}^\top R_{P}^{\dagger} S_{P} &= 0,
	\end{align*}
	such that Assumption~\ref{ass:dissipativity} holds with $\Lambda=P$. $\hfill\qed$
\end{pf}
We observe that the condition $R_P\succeq0$ is necessary for the optimal control input yielded by~\eqref{eq:ocp} to remain bounded, as also discussed in~\cite{Stoorvogel1998}. This proves the necessity of quadratic pre-dissipativity for the OCP~\eqref{eq:ocp} to have bounded solutions $\|u(x)\| < \infty$ for all $\|x\|<\infty$, $\|v\|<\infty$.

\subsection{Exponential Stability for Controllable $(A,B)$}

In this section, we discuss the stability properties of the closed-loop system~\eqref{eq:optimal_closed_loop}, defined as follows. 
\begin{Definition}[Stability]
	Given a system $x_+=Ax$, the origin is
	\begin{enumerate}
		\item[(i)] \emph{stable} if for every $\epsilon>0$ there exists $\delta(\epsilon)>0$ such that $\|A^kx\|<\epsilon$ for all $k$ and all $\|x\|<\delta(\epsilon)$;
		\item[(ii)] \emph{(globally) asymptotically stable} if it is stable and $\lim_{k\to\infty}A^k x=0$
	\end{enumerate}
\end{Definition}
We observe that~\cite[Theorem~10.9]{Antsaklis1997}:
\begin{enumerate}
	\item[(i)] stability $\Leftrightarrow$ all eigenvalues of $A$ are within or on the unit circle of the complex plane, and every eigenvalue that is on the unit circle has an associated Jordan block of order $1$;
	\item[(ii)] asymptotic stability $\Leftrightarrow$ all eigenvalues of $A$ are within the unit circle, i.e., matrix $A$ is Schur stable.
\end{enumerate}
Consequently, asymptotic stability is a global property and is equivalent to exponential stability.

Conditions guaranteeing exponential stability of the closed-loop system~\eqref{eq:optimal_closed_loop} have been claimed in \cite{Olanrewaju2017}, and rely on the following assumption.

\begin{Assumption}
	\label{ass:stability_rank}
	There exist symmetric matrices $\Lambda_1$, $\Lambda_2$ such that 
	\begin{align*}
		H_{\Lambda_1}
		&\succeq 0, &
		H_{\Lambda_2}
		&\succeq 0, \\
		\mathrm{rank}(\Lambda_1 - \Lambda_2) &=n_x.
	\end{align*}
\end{Assumption}
We will prove next that the following seemingly stronger assumptions are both necessary and sufficient for asymptotic stability. 

\begin{Assumption}
	\label{ass:stability_pd}
There exist symmetric matrices $\Lambda_1$, $\Lambda_2$ such that 
\begin{align*}
	H_{\Lambda_1}
	&\succeq 0, &
	H_{\Lambda_2}
	&\succeq 0, \\
	\Lambda_1 - \Lambda_2 &\succ 0.
\end{align*}
\end{Assumption}

\begin{Assumption}[Quadratic Strict Pre-Dissipativity]
	\label{ass:strict_dissipativity}
	There exists a symmetric matrix $\Lambda$ such that 
	\begin{align}
		\label{eq:strict_dissipativity}
		H_{\Lambda} &\succeq 0, &
		Q_{\Lambda} -S_{\Lambda}^\top R_{\Lambda}^{\dagger} S_{\Lambda} &\succ 0.
	\end{align}
\end{Assumption}

Before proving our results, we observe that Assumption~\ref{ass:strict_dissipativity} is the linear-quadratic equivalent of the standard strict pre-dissipativity assumption for the nonlinear case, i.e., Definition~\ref{def:nl_strict_diss} where $\mathcal{H}$ need not be compact~\cite{Gruene2021}. 
As we will prove that Assumptions~\ref{ass:stability_pd},~\ref{ass:strict_dissipativity} are equivalent, we will refer to both of them as quadratic strict pre-dissipativity whenever the distinction is not relevant. We will discuss in Section~\ref{sec:checking_assumption} how these assumptions can be checked.

In~\cite{Olanrewaju2017} it is claimed that Assumption~\ref{ass:stability_rank} is sufficient and necessary for exponential stability. We prove next that this holds for Assumptions~\ref{ass:stability_pd} and~\ref{ass:strict_dissipativity}.
\begin{Theorem}
	\label{thm:stability}
	Suppose that $(A,B)$ is controllable and either Assumption~\ref{ass:stability_pd} or Assumption~\ref{ass:strict_dissipativity} holds. Then, the CGDARE has a stabilizing solution, such that the closed-loop system~\eqref{eq:optimal_closed_loop} is exponentially stable and $BG=0$. Moreover, both Assumption~\ref{ass:stability_pd} and~\ref{ass:strict_dissipativity} are necessary for exponential stability, hence they imply each other.
\end{Theorem}
\begin{pf}
	We consider first the case in which Assumption~\ref{ass:strict_dissipativity} holds for some matrix $\Lambda_2$. This immediately entails that a full-rank solution $P_{\Lambda_2}\succ0$ exists, such that, by Lemma~\ref{lem:BG0} we have $BG=0$. One can prove that  $P_{\Lambda_2}\succ0$ by observing that, starting with zero terminal cost, the first step of the Riccati iterations (i.e., value iteration) yields cost-to-go $P_{\Lambda_2,1}=Q_{\Lambda_2} -S_{\Lambda_2}^\top R_{\Lambda_2}^{\dagger} S_{\Lambda_2} \succ0$, such that for all $n>1$ we have $P_{\Lambda_2}\succeq P_n \succeq P_1 \succ0$~\cite[Lemma~3.17]{Zanon2025}. Consequently, we can alternatively rotate the cost using $\Lambda_1 = \Lambda_2 + P_{\Lambda_2}\succ \Lambda_2$, which yields a positive semidefinite stage cost matrix, such that Assumption~\ref{ass:strict_dissipativity} entails Assumption~\ref{ass:stability_pd}.
	
	We prove next that Assumption~\ref{ass:stability_pd} yields exponential stability.
	We define $\Lambda_{1,2}:=\Lambda_1-\Lambda_2 \succ0$, such that
	\begin{align*}
		H_{\Lambda_1} &= H_{\Lambda_2+\Lambda_{1,2}}, 
	\end{align*}
	By Theorem~\ref{thm:existence} $H_{\Lambda_1}\succeq0$ entails the existence of a solution $P_{\Lambda_1}\succeq0$ and, in turn of a solution
	\begin{align*}
		P_{\Lambda_2} &= P_{\Lambda_1} + \Lambda_{1,2} \succ0,
	\end{align*}
	which, by Lemma~\ref{lem:BG0} entails $BG=0$.
	Moreover, by defining $K=R_{P_{\Lambda_2}}^\dagger S_{P_{\Lambda_2}}$, we have
	\begin{align*}
		&(A-BK)^\top P_{\Lambda_2} (A-BK)  - P_{\Lambda_2} \\
		&\hspace{3em}=-\left( Q_{\Lambda_2} - K^\top S_{\Lambda_2} - S_{\Lambda_2}^\top K + K^\top R_{\Lambda_2} K\right) \preceq 0.
	\end{align*}
	This entails that $A-B K$ has all eigenvalues inside the unit circle, possibly except for some of them on the unit circle (corresponding to 1 by 1 Jordan blocks)~\cite[Theorem~10.11]{Antsaklis1997}. 
	
	Because the system is controllable, an exponentially stabilizing solution to OCP~\eqref{eq:ocp} formulated with the additional constraint $\lim_{k\to\infty} x_k =0$ exists and yields a linear feedback $K_\mathrm{s}$ with quadratic cost-to-go matrix $P_\mathrm{s}$. This proves sufficiency.
	
	In order to prove necessity, consider the case in which one of the solutions $P_\mathrm{s}$, $K_\mathrm{s}$ of the CGDARE yields exponential stability, such that $BG=0$. Because $P_\mathrm{s}$ can be freely chosen by a suitable cost rotation, we will assume without loss of generality that $P_\mathrm{s}\succ0$.
	Moreover, because $K_\mathrm{s}$ yields exponential stability, for any $\tilde Q \succ 0$ there exists a matrix $V\succ0$ satisfying
	\begin{align*}
		(A-BK_\mathrm{s})^\top V (A-BK_\mathrm{s}) - V = - \tilde Q\prec 0.
	\end{align*}
	Let us select $\Lambda=P_\mathrm{s}-V$ such that $P_{\Lambda,s}= V \succ0$, 
	and, because $BG=0$, $H_{\Lambda}\succeq0$ if and only if
	\begin{align*}
		R_{\Lambda} &\succeq 0, &
		Q_{\Lambda} - S_{\Lambda}^\top R_{\Lambda}^\dagger S_{\Lambda} &\succeq0.
	\end{align*}
	The first condition holds for $\tilde Q$ small enough because $R_{P_\mathrm{s}}\succeq0$, and $P_\mathrm{s}\succ0$, while the second one can be transformed as follows. 
	We rewrite the rotated CGDARE as
	\begin{align*}
		&(A-BK_\mathrm{s})^\top P_{\Lambda,s} (A-BK_\mathrm{s}) - P_{\Lambda,s} \\
		&\hspace{2em}= -(Q_{\Lambda} - K_\mathrm{s}^\top S_{\Lambda} - S_{\Lambda}^\top K_\mathrm{s} + K_\mathrm{s}^\top R_{\Lambda} K_\mathrm{s})=-\tilde Q,
	\end{align*}
	such that we can write
	\begin{align*}
		&\hspace{-1em}Q_{\Lambda} - S_{\Lambda}^\top R_{\Lambda}^\dagger S_{\Lambda} \\
		&=\tilde Q - S_{\Lambda}^\top R_{\Lambda}^\dagger S_{\Lambda} + S_{\Lambda}^\top K_\mathrm{s} + K_\mathrm{s}^\top S_{\Lambda} - K_\mathrm{s}^\top R_{\Lambda} K_\mathrm{s}  \\
		&=\tilde Q - (S_{\Lambda}^\top - K_\mathrm{s}^\top R_{\Lambda}) R_{\Lambda}^\dagger (S_{\Lambda} - R_{\Lambda} K_\mathrm{s}  ).
	\end{align*}
	We observe next that, because $K_\mathrm{s}=R_{P_\mathrm{s}}^\dagger S_{P_\mathrm{s}}$,
	\begin{align*}
		S_{\Lambda} - R_{\Lambda} K_\mathrm{s} &= S_{P_\mathrm{s}} - R_{P_\mathrm{s}}K_\mathrm{s}  - B^\top V (A - B K_\mathrm{s}), \\
		&= -B^\top V (A - B K_\mathrm{s}).
	\end{align*}
	Consequently, 
	\begin{align*}
		&\tilde Q -  (S_{\Lambda}^\top - K_\mathrm{s}^\top R_{\Lambda}) R_{\Lambda}^\dagger (S_{\Lambda} - R_{\Lambda} K_\mathrm{s}  ) \\
		&\hspace{1em}=\tilde Q-(A-B{K_\mathrm{s}})^\top V B R_{\Lambda}^\dagger B^\top V (A - B{K_\mathrm{s}}  ).
	\end{align*}
	Therefore, $H_{\Lambda}\succeq0$ if and only if
	\begin{align}
		\label{eq:lyap_diss_condition}
		R_{\Lambda}&\succeq0, & \tilde Q &\succeq (A - B_{K\mathrm{s}}  )^\top \tilde V (A - B{K_\mathrm{s}}  ),
	\end{align}
	where $\tilde V:= V B R_{\Lambda}^\dagger B^\top V$. 
	We prove next that Condition~\eqref{eq:lyap_diss_condition} can be satisfied for a sufficiently small $\tilde Q$. Assume that the condition does not hold for some $\hat Q\succ 0$ yielding Lyapunov matrix $\hat V$, and select $\tilde Q = a\hat Q$ with $a>0$. Note that for $a\leq 1$ we have $0\preceq R_{\Lambda}=R_{P_\mathrm{s}-V}=R_{P_\mathrm{s}-a\hat V}\succeq R_{P_\mathrm{s}-\hat V}$, such that $R_{\Lambda}^\dagger \preceq R_{P_\mathrm{s}-\hat V}^\dagger$ and
	\begin{align*}
		\tilde V &\preceq V B R_{P_\mathrm{s}-\hat V}^\dagger B^\top V =a^2 \hat V B R_{P_\mathrm{s}-\hat V}^\dagger B^\top \hat V,
	\end{align*}
	i.e., $\tilde V$ is upper bounded by a quantity that scales quadratically with $a$, while $\tilde Q$ scales linearly. Hence, for $a$ sufficiently small Condition~\eqref{eq:lyap_diss_condition} holds. Moreover, also $R_{\Lambda}\succeq0$ holds for $a$ sufficiently small. 
	This also proves that in~\eqref{eq:lyap_diss_condition} the second inequality can be enforced in a strict sense, therefore proving necessity of Assumption~\ref{ass:strict_dissipativity} for exponential stability. 
	As we proved that Assumption~\ref{ass:strict_dissipativity} implies Assumption~\ref{ass:stability_pd} this concludes the proof. 
	$\hfill\qed$
\end{pf}

The result of Theorem~\ref{thm:stability} entails that the closed-loop system is asymptotically stable for all optimal control inputs. However, there are also cases in which some optimal control inputs stabilize the system, while others do not, as shown by the following example. 
While an in-depth analysis of such cases is beyond the scope of this paper, we will provide a brief discussion in Section~\ref{sec:difficult_case}.

\begin{Example}
	\label{ex:possibly_destabilizing}
	Consider the following system:
	\begin{align*}
		A &= \matr{ll}{0.9 & 1 \\ 0 & 1}, & B &= \matr{rr}{2 & 0 \\ 1 & 1},\\
		Q &= \matr{ll}{0 & 0 \\ 0 & 1}, & R &= 0, &S &= 0.
	\end{align*}
	The solution of the CGDARE is
	\begin{align*}
		P &= \matr{rr}{0 & 0 \\ 0 & 1}, & K &= \matr{rr}{0 & 0.5 \\ 0 & 0.5}, & G &= \matr{rr}{0.5 & -0.5 \\ -0.5 & 0.5},
	\end{align*}
	such that 
	\begin{align*}
		A-BK = \matr{ll}{0.9 & 0 \\ 0 & 0}, 
	\end{align*} has all eigenvalues strictly inside the unit circle, but some optimal solutions can destabilize the system, e.g., for feedback matrix $L=-I$, yielding $v=-Lx$ we have
	\begin{align*}
		A-BK - BGL = \matr{lr}{1.9 & -1 \\ 0 & 0}, 
	\end{align*}
	which has one unstable eigenvalue.
	Note that, while Assumption~\ref{ass:dissipativity} holds, Assumptions~\ref{ass:stability_pd},~\ref{ass:strict_dissipativity} do not hold. Indeed, as stated by Theorems~\ref{thm:existence} and~\ref{thm:stability}, a positive semi-definite solution $P$ yielding the optimal cost-to-go exists, but the closed-loop system is not exponentially stable for all optimal control inputs.
	
	Note that, if
	\begin{align*}
		A &= \matr{ll}{1 & 1 \\ 0 & 1},
	\end{align*}
	similar considerations follow, but $v=0$ does not yield exponential stability, even though there exist infinitely many matrices $L$ such that $v=-Lx$ does yield exponential stability. See~\cite[Example~5.1]{Ferrante2013} for more details.
\end{Example}

This example shows that, whenever $BG\neq0$, there exist optimal control inputs that destabilize the system, while other optimal control inputs might stabilize the system. Indeed, if $BG\neq0$ Assumptions~\ref{ass:stability_pd}, and~\ref{ass:strict_dissipativity} cannot hold, as they do imply $BG=0$.

\subsection{Exponential Stability for Stabilizable $(A,B)$}
The results above focused on the case in which $(A,B)$ is controllable. We extend them next to the case in which $(A,B)$ is stabilizable. 

\begin{Theorem}
	\label{thm:as_stab_no_ctrb}
	Suppose that $(A,B)$ is stabilizable. Then, Assumptions~\ref{ass:stability_pd} and~\ref{ass:strict_dissipativity} imply each other. If any of them holds, the closed-loop system~\eqref{eq:optimal_closed_loop} is exponentially stable and $BG=0$. Moreover, Assumptions~\ref{ass:stability_pd}, and~\ref{ass:strict_dissipativity} are also necessary for exponential stability.
\end{Theorem}
\begin{pf}
	Let us assume without loss of generality that the system is in the form of the Kalman controllability decomposition
	\begin{align}
		\label{eq:Kalman_form}
		A&=\matr{rr}{A_{11} & 0\\A_{21} & A_{22}}, & B&=\matr{c}{0\\ B_2},
	\end{align}
	with $(A_{22},B_2)$ controllable, and $A_{11}$ asymptotically stable.
	Then we have
	\begin{align*}
		B^\top P B &= B_2^\top P_{22} B_2; \\
		B^\top P A 
		&= \matr{cc}{ s_1(P_{12},P_{22}) & s_2(P_{22}) },
	\end{align*}
	with 
	\begin{align*}
		s_1(P_{12},P_{22})&:= B_2^\top P_{12}^\top A_{11} + B_2^\top P_{22} A_{21}, \\
		s_2(P_{22}) &:= B_2^\top P_{22} A_{22};
	\end{align*}
	and
	\begin{align*}
		A^\top P A &= \matr{cc}{ q_{11}(P_{11},P_{12},P_{22}) & q_{12}(P_{12},P_{22}) \\ q_{12}(P_{12},P_{22})^\top & q_{22}(P_{22}) }, 
	\end{align*}
	with
	\begin{align*}
		q_{11}(P_{11},P_{12},P_{22}) &:= A_{11}^\top P_{11}A_{11} + A_{11}^\top P_{12}A_{21} \\
		&\hspace{4em}+ A_{21}^\top P_{12}^\top A_{11} + A_{21}^\top P_{22} A_{21}, \\
		q_{12}(P_{12},P_{22}) &:= A_{11}^\top P_{12}A_{22} + A_{21}^\top P_{22} A_{22}, \\
		q_{22}(P_{22}) &:= A_{22}^\top P_{22} A_{22}.
	\end{align*}
	Consequently,
	\begin{align*}
		H_P &= \matr{cc}{ H_P^\mathrm{u} & {H_P^\mathrm{m}}^\top \\ H_P^\mathrm{m} & H_{P_{22}}^\mathrm{c} },\\
		H_P^\mathrm{u} &:= Q_{11} + q_{11}(P_{11},P_{12},P_{22}) - P_{11}, \\
		H_P^\mathrm{m} &:= \matr{c}{ Q_{21} \!+\! q_{12}(P_{12},P_{22})^\top - P_{12}^\top  \\  S_1 \!+\! s_1(P_{12},P_{22}) }, \\
		H_{P_{22}}^\mathrm{c} &:= \matr{cc}{Q_{22} + q_{22}(P_{22}) - P_{22} & S_2^\top + s_2(P_{22})^\top \\  S_2 + s_2(P_{22}) & R + B_2^\top P_{22} B_2 }.
	\end{align*}
	This entails that the lower right block $H_{P_{22}}^\mathrm{c}$ only depends on $P_{22}$, i.e., the controllable subsystem is independent of the rest. We can then apply Theorem~\ref{thm:stability} to the controllable subsystem.
	
	We prove next that, if quadratic strict pre-dissipativity holds for the controllable subsystem, then it also holds for the full system, since all uncontrollable modes are exponentially stable. Note that, by Theorem~\ref{thm:stability} the two quadratic strict pre-dissipativity assumptions~\ref{ass:stability_pd},~\ref{ass:strict_dissipativity} are equivalent for controllable systems, hence we do not need to distinguish among them.
	
	By Lemma~\ref{lem:BG0} quadratic strict pre-dissipativity entails $B_2G=0$ and, consequently, $BG=0$. Hence,
	\begin{align*}
		H_{P_{22}}^{\mathrm{c},G} &:= H_{P_{22}}^{\mathrm{c}} + \matr{cc}{0 & 0\\  0 & G } \succ 0.
	\end{align*}
	Moreover, stabilizability entails that $A_{11}$ has all eigenvalues inside the unit circle. Hence, for fixed matrices $P_{12}$, $P_{22}$, the quantity $A_{11}^\top P_{11}A_{11} - P_{11}$ can be made arbitrarily large by a suitable choice of $P_{11}$. Consequently, if we rewrite the blocks of $H_P$ such that
	\begin{align*}
		H_P &= \matr{cc}{ \Omega & \Sigma^\top \\ \Sigma & \Theta},
	\end{align*}
	where $\Theta := R + B_2^\top P_{22} B_2\succeq0$, we have that
	\begin{align*}
		\Omega - \Sigma^\top \Theta^{\dagger} \Sigma &= \Omega - \Sigma^\top (\Theta+G)^{\dagger} \Sigma \\
		&= \Omega - \Sigma^\top (\Theta+G)^{-1} \Sigma \succ 0,
	\end{align*}
	i.e., Assumption~\ref{ass:strict_dissipativity} holds for the full system for some matrix $\Lambda$. 
	Choose $\Lambda_2=\Lambda$ and $\Lambda_{1,2} =  \Omega - \Sigma^\top \Theta^{\dagger} \Sigma$. Then
	\begin{align*}
		H_{\Lambda_1} &= H_{\Lambda_2} + \matr{cc}{A^\top \Lambda_{1,2} A - \Lambda_{1,2} & A^\top \Lambda_{1,2} B \\
		B^\top \Lambda_{1,2} A & B^\top \Lambda_{1,2} B} \\
		&\succeq H_{\Lambda_2} + \matr{cc}{- \Lambda_{1,2} & 0 \\
			0 & 0} \succeq 0,
	\end{align*}
	i.e., Assumption~\ref{ass:stability_pd} holds for the full system.
	$\hfill\qed$
\end{pf}

\begin{Remark}
	Lemma~\ref{lem:BG0_rdare} states that if $BG=0$ the solutions of the CGDARE are solutions of the rDARE and vice versa, such that the CGDARE inherits also all well-known properties of DAREs: if it exists, the stabilizing solution is unique, and all solutions $P$ satisfy $P_\mathrm{s} \succeq P$, if $P_\mathrm{s}$ exists; $P\succeq\bar P_\mathrm{s}$, if $\bar P_\mathrm{s}$ exists; $P_\mathrm{s} \succ \bar P_\mathrm{s}$ if both exist.
\end{Remark}

\subsection{Checking Quadratic Strict Pre-Dissipativity}
\label{sec:checking_assumption}

While the theory we developed proved that the quadratic strict pre-dissipativity assumptions we considered are equivalent, we did not yet comment how they can be checked in practice. 
A procedure to check Assumption~\ref{ass:stability_rank} in several steps is given in~\cite{Olanrewaju2017}. Assumption~\ref{ass:strict_dissipativity} is not easy to check directly as the second inequality is nonlinear, though we will discuss later one possible reformulation. Assumption~\ref{ass:stability_pd}, can be checked directly by solving one suitably formulated SDP.
One option consists in solving
\begin{subequations}
	\label{eq:diss_sdp2}
	\begin{align}
		\min_{\Lambda_1,\Lambda_2,a} \ & \! -a \\
		\mathrm{s.t.} \ \ & H_{\Lambda_1} \succeq 0, \\
		& H_{\Lambda_2} \succeq 0, \\
		& \Lambda_1 - \Lambda_2 \succeq aI,
	\end{align}
\end{subequations}
and checking if $a>0$.

We provide next an alternative formulation, which we consider insightful for the case in which $(A,B)$ is controllable.
As an alternative to~\eqref{eq:diss_sdp2}, one can solve
\begin{subequations}
	\label{eq:diss_sdp}
	\begin{align}
		\min_{\Lambda_1,\Lambda_2} \ & \! -\mathrm{tr}(\Lambda_1-\Lambda_2) \\
		\mathrm{s.t.} \ \ & H_{\Lambda_1} \succeq 0, \\
		& H_{\Lambda_2} \succeq 0.
	\end{align}
\end{subequations}
and check whether $\Lambda_1\succ \Lambda_2$.

For controllable systems SDP~\eqref{eq:diss_sdp} allows us to further discuss the tight relation between quadratic strict pre-dissipativity, especially in the form of Assumption~\ref{ass:stability_pd}, and the stabilizing solutions of the CGDARE and RCGDARE. In fact, as we prove in the next lemma, if quadratic strict pre-dissipativity holds and $(A,B)$ is controllable, its solution is $\Lambda_1=P_\mathrm{s}$, $\Lambda_2=\bar P_\mathrm{s}$.
\begin{Lemma}
	Assume that $(A,B)$ is controllable. Then, any $\Lambda$ such that $H_\Lambda\succeq0$ satisfies $P_\mathrm{s} \succeq \Lambda$ if $P_\mathrm{s}$ exists; and $\Lambda\succeq \bar P_\mathrm{s}$ if $\bar P_\mathrm{s}$ exists. 
\end{Lemma}
\begin{pf}
		Because $H_\Lambda\succeq0$ entails the existence of $P_\Lambda\succeq0$, we have $P_\mathrm{s}\succeq P=P_\Lambda+\Lambda\succeq\Lambda$. The RCGDARE instead is such that $H_\Lambda\succeq0$ entails the existence of $\bar P_\Lambda\preceq0$, which yields $\bar P_\mathrm{s}\preceq \bar P= \bar P_\Lambda+\Lambda\preceq\Lambda$.
$\hfill\qed$
\end{pf}

In the lemma above, we assumed controllability, because if the uncontrollable modes in $A$ are stable, then the uncontrollable modes in $\bar A$ are unstable. Consequently, $\bar P_\mathrm{s}$ does not exist. 
In case the system is not controllable, assuming without loss of generality that the system is in the Kalman controllability decomposition~\eqref{eq:Kalman_form}, one can partition matrix $E$ consistently as 
\begin{align*}
	E=\matr{cc}{E_{11} & E_{12}\\E_{21}&E_{22}},
\end{align*}
where $E_{11}$ and $E_{12}=E_{21}^\top$ can be chosen arbitrarily, while $E_{22}\preceq0$.

For $E_{12}=E_{21}^\top=0$, $E_{22}=0$ this yields
\begin{align*}
	S_{P_\mathrm{s}+E}^\top R_{P_\mathrm{s}+E}^{\dagger} S_{P_\mathrm{s}+E} = S_{P_\mathrm{s}}^\top R_{P_\mathrm{s}}^{\dagger} S_{P_\mathrm{s}}.
\end{align*}
Additionally, $E_{11}$ can be chosen such that
\begin{align*}
	A_{11}^\top E_{11} A_{11} - E_{11} \prec 0.
\end{align*}
Consequently, any matrix satisfying the inequality above can be multiplied by an arbitrarily large constant while still satisfying the inequality. Hence, the SDP solution is unbounded. 
Clearly, infinitely many others are also possible. Consequently, if $(A,B)$ is stabilizable but not controllable, SDP~\eqref{eq:diss_sdp} is unbounded. A simple fix can then be to introduce a constraint of the form
\begin{align*}
	\Lambda_1 - \Lambda_2 \preceq bI,
\end{align*}
for some fixed and sufficiently large constant $b$. Alternatively, one could add the term $b(\mathrm{tr}(\Lambda_1-\Lambda_2))^2$ to the cost for some fixed constant $b$.

Finally, we would like to stress that, though Assumption~\ref{ass:strict_dissipativity} is formulated as a nonlinear matrix inequality, one can check if the assumption holds by one additional indirect approach, which rewrites the condition by exploiting the properties of the Schur complement.
\begin{Lemma}
	Strict dissipativity Condition~\eqref{eq:strict_dissipativity} is equivalent to
	\begin{align}
		\label{eq:strict_dissipativity_smart}
		H_\Lambda^a := \matr{c@{\hspace{11pt}}c}{Q_{\Lambda} - aI & S_{\Lambda}^\top \\ S_{\Lambda} & R_{\Lambda}} &\succeq 0, & a &> 0.
	\end{align}
\end{Lemma}
\begin{pf}
	As proven in~\cite[Lemma~2.1]{Ferrante2004}, see also~\cite[Lemma~2.1]{Ferrante2013}, if~\eqref{eq:strict_dissipativity_smart} holds, then 
	\begin{align*}
		&S_\Lambda^\top G_\Lambda = S_\Lambda^\top (I-R_\Lambda^\dagger R_\Lambda) =0, \\ 
		&Q_{\Lambda} - aI -S_{\Lambda}^\top R_{\Lambda}^{\dagger} S_{\Lambda} \succeq 0,
	\end{align*}
	such that we immediately obtain that~\eqref{eq:strict_dissipativity_smart} entails~\eqref{eq:strict_dissipativity}. Consequently, strict dissipativity must hold. In order to prove the converse, we observe that if strict dissipativity holds, because $R_\Lambda\succeq0\implies R_\Lambda+G_\Lambda\succ0$, we immediately obtain also
	\begin{align}
		\label{eq:strict_dissipativity_reg}
		\matr{c@{\hspace{11pt}}c}{Q_{\Lambda} & S_{\Lambda}^\top \\ S_{\Lambda} & R_{\Lambda}+G_\Lambda} &\succ 0.
	\end{align}
	Additionally, because $BG=0$, we have that $G$ is independent of $\Lambda$, i.e., $G=G_\Lambda$, and $S_\Lambda^\top G_\Lambda =0$ and $R_\Lambda G_\Lambda =0$. Consequently,~\eqref{eq:strict_dissipativity_smart} and~\eqref{eq:strict_dissipativity_reg} are equivalent conditions, which, in turn entails that~\eqref{eq:strict_dissipativity_smart} and~\eqref{eq:strict_dissipativity} are equivalent. 
	$\hfill\qed$
\end{pf}
This lemma proves that strict dissipativity is equivalent to dissipativity for a modified stage cost, such that one can check strict dissipativity by solving the SDP
\begin{align*}
	\max_{a,\Lambda} & \ \ a & \mathrm{s.t.} &          \ \ H_\Lambda^a \succeq0,
\end{align*}
and checking if the optimal solution satisfies $a>0$. 

\section{Finite Horizon and Terminal Cost}
\label{sec:terminal_cost}

In this section, we consider the finite-horizon case and we discuss the role of the terminal cost in providing exponential stability. To that end, we exploit the theory developed so far to extend the results of~\cite{Zanon2025} to the case in which quadratic strict $(x,u)$ pre-dissipativity does not hold, but quadratic strict pre-dissipativity does hold. The main question is how exponential stability can be enforced in finite-horizon EMPC by means of suitable terminal costs, without terminal constraints.

We introduce the Receding-Horizon OCP (RHOCP)
\begin{subequations}
	\label{eq:rh-ocp}
	\begin{align}
		V(\hat x_0) = \min_{x,u} \ & \sum_{k=0}^{N-1} \matr{c}{x_k \\ u_k}^\top H \matr{c}{x_k \\ u_k} + x_N^\top P^\mathrm{f} x_N \\
		\mathrm{s.t.} \ & x_0 = \hat x_0, \\
		&x_{k+1} = A x_k + B u_k,
	\end{align}
\end{subequations}
which delivers the optimal feedback law
\begin{align}
	\label{eq:rh_feedback}
	F_N(x,v) = -K_Nx + G_Nv, && v \text{ arbitrary},
\end{align}
with $P_N, K_N$ defined by the generalized Riccati recursion
\begin{subequations}
	\label{eq:gen_riccati_recursion}	
	\begin{align}
		P_{N+1} =\ & Q + A^\top P_N A - (S^\top + A^\top P_N B)K_{N+1}, \\
		K_{N+1} =\ & (R+B^\top P_N B)^{\dagger} (S+B^\top P_N A),
	\end{align}
\end{subequations}
where $P_0=P^\mathrm{f}$, and
\begin{align*}
	G_N := I - R_{P_{N-1}}^\dagger R_{P_{N-1}}.
\end{align*}
The corresponding closed-loop system is
\begin{align}
	\label{eq:optimal_closed_loop_N}
	x_+ = (A-BK_N)x + BG_Nv, && v \text{ arbitrary}.
\end{align}

In order to obtain the desired results, we will exploit the fact that, if quadratic strict pre-dissipativity holds, then the properties of the CGDARE can be studied by exploiting the rDARE instead. Moreover, if $BG=0$, then (see Lemma~\ref{lem:BG0_rdare})
\begin{align*}
	R_\Lambda G&= 0, & S_\Lambda^\top G&=0,
\end{align*}
holds for all $\Lambda$, and not just for those solving the CGDARE. This will be particularly helpful  for the finite horizon case, as the finite-horizon solution does in general not match the infinite-horizon solution. 
Additionally, also the RCGDARE can be studied by exploiting the corresponding rRDARE, as we prove in the next theorem.

\begin{Theorem}[{Extension of~\cite[Theorem~4.7]{Zanon2025}}]
	Consider a strictly pre-dissipative OCP~\eqref{eq:ocp} with controllable $(A,B)$ and let $\bar P_\mathrm{s}$ be the stabilizing solution of the RCGDARE. Choose the terminal cost matrix $P^\mathrm{f} \succ \bar P_\mathrm{s}$. Then the optimal solution is given by the unique stabilizing solution of the corresponding rDARE. In particular, the origin is an asymptotically stable equilibrium of the optimally controlled system.
\end{Theorem}
\begin{pf}
	The key observation is given in Lemma~\ref{lem:BG0_rdare}, which states that, as quadratic strict pre-dissipativity entails $BG=0$ and $R_P\succeq0$ for all $P$ solving the CGDARE, then the solutions of the CGDARE and the rDARE coincide. Consequently, we can directly apply~\cite[Theorem~4.7]{Zanon2025} to the rDARE. $\hfill\qed$
\end{pf}

\begin{Theorem}[{Extension of~\cite[Theorem~4.9]{Zanon2025}}]
	Consider a controllable pair $(A,B)$ equipped with a strictly pre-dissipative stage cost and let $\bar P_\mathrm{s}$ be the stabilizing solution of the corresponding RCGDARE. Select the terminal cost matrix as $P^\mathrm{f} \succ \bar P_\mathrm{s}$. Then, for any sufficiently large finite horizon $N$ the RHOCP~\eqref{eq:rh-ocp} yields a closed-loop system~\eqref{eq:optimal_closed_loop_N} for which the origin is globally exponentially stable.
\end{Theorem}
\begin{pf}
	Also in this case, we exploit Lemma~\ref{lem:BG0_rdare}, in particular the fact that~\eqref{eq:rDARE_CGDARE} holds for any $\Lambda$, i.e., $G_N=G$, and the equivalence between the original formulation and the one with regularization $u^\top G u$ holds not only for an infinite horizon (i.e., the CGDARE and rDARE share the same solutions), but also throughout the generalized Riccati iterations (i.e., value iteration). We can then  apply~\cite[Theorem~4.9]{Zanon2025} to the regularized problem and carry over the result to the original problem. $\hfill\qed$
\end{pf}

The extension of the results above to the stabilizable case is directly obtained by applying the arguments used in Theorem~\ref{thm:as_stab_no_ctrb} and noticing that the choice of bounded $P^\mathrm{f}_{11}, P^\mathrm{f}_{12}$ can be made arbitrarily (as long as it is bounded), as it does not affect the feedback on the controllable modes and only shifts the non-controllable blocks of $P_N$.

\section{The Stabilizable but Possibly Unstable Case}
\label{sec:difficult_case}

In this section, we briefly discuss the case in which the CGDARE does have a solution $P$ satisfying $R+B^\top P B \succeq 0$, but does not yield an exponentially stable optimal closed-loop system~\eqref{eq:optimal_closed_loop}. This case is discussed in~\cite{Ferrante2013} and in Example~\ref{ex:possibly_destabilizing}, which is taken from~\cite{Ferrante2013}. 

Note that, if $BG=0$, by Lemma~\ref{lem:BG0_rdare} all $P,K$ solutions of the CGDARE are also solutions of the rDARE and vice versa. Hence, quadratic strict $(x,u)$ pre-dissipativity cannot hold for the rDARE and, in turn, quadratic strict pre-dissipativity cannot hold for the CGDARE. 
When instead, $BG\neq 0$, the optimal closed-loop system~\eqref{eq:optimal_closed_loop} does not define a single state trajectory, and the cost might allow both stabilizing and destabilizing optimal solutions. In case some of the optimal solutions are stabilizing, one might be interested in selecting one of those. Unfortunately, straightforward attempts at generalizing quadratic strict pre-dissipativity yield inequalities of the form
\begin{align*}
	H_\Lambda + \matr{cc}{L^\top T L & L^\top T \\ TL & T } &\succeq 0, & \matr{c}{PB \\ R}T &=0.
\end{align*}
Note that the first condition is a nonlinear matrix inequality and the second condition is not easy to satisfy without solving the CGDARE first. However, if one does solve the CGDARE, one can pick $T=G$, but the first condition still remains nonlinear in $L$. 
The idea to remove nonlinearities by regularizing the control input only, i.e., by selecting $L=0$, cannot work if some uncontrollable and unstable mode has zero cost. 
Such a situation occurs, e.g., in the second case in Example~\ref{ex:possibly_destabilizing}, as we explain next.

\begin{Example}
	Consider once more the second system in Example~\ref{ex:possibly_destabilizing}, i.e.:
	\begin{align*}
		A &= \matr{ll}{1 & 1 \\ 0 & 1}, & B &= \matr{rr}{2 & 0 \\ 1 & 1},\\
		Q &= \matr{ll}{0 & 0 \\ 0 & 1}, & R &= 0, &S &= 0.
	\end{align*}
	Any cost rotation yields
	\begin{align*}
		Q + A^\top \Lambda A - \Lambda = \matr{cc}{0 & \Lambda_{11} \\ \Lambda_{11} & 1 + \Lambda_{11} + 2\Lambda_{12}} \nsucc 0.
	\end{align*}
	Consequently, there exists no cost rotation that can satisfy Assumption~\ref{ass:strict_dissipativity}.
	
	However, if the system is pre-stabilized first, e.g., using feedback $\hat K=0.5I$, $A-B\hat K$ has all eigenvalues inside the unit circle, and in this specific case the cost matrices remain unchanged~\cite{Zanon2025}, since $S=0$, $R=0$. 
	A solution of the CGDARE is
	\begin{align*}
			P &= \matr{rr}{0 & 0 \\ 0 & 1}, & K &= \matr{rr}{-0.25 & 0.25 \\ -0.25 & 0.25}, & G &= \matr{rr}{0.5 & -0.5 \\ -0.5 & 0.5}.
	\end{align*}
	As expected, $P$ is unchanged, while $K$ does change. In this case, regularizing the rotated cost as
	\begin{align*}
		H_\Lambda^G := H_\Lambda + \matr{cc}{0 & 0 \\ 0 & G },
	\end{align*}
	makes it possible to find a matrix $\Lambda$ such that $H_\Lambda^G\succ0.$
\end{Example}

The example above suggests a possible, though a bit cumbersome, procedure to tackle the case in which some optimal controls might be stabilizing while others might not be. In particular, one can: (i) pre-stabilize the system using the desired technique to compute feedback matrix $\hat K$ and update the cost matrices as
\begin{align*}
	\hat H &= \matr{ll}{\hat Q & \hat S^\top \\ \hat S & R}, \\
	\hat Q &:= Q - S^\top {\hat K} - {\hat K}^\top S + {\hat K}^\top R {\hat K}, \\
	\hat S &:= S - R{\hat K};
\end{align*}
(ii) compute any solution of the CGDARE to obtain $G$ and define
\begin{align*}
	\hat H_\Lambda^G := \hat H_\Lambda + \matr{cc}{0 & 0 \\ 0 & G};
\end{align*}
(iii) define the cost of the corresponding rDARE and verify if quadratic strict $(x,u)$ dissipativity holds, i.e., if there exists some matrix $\Lambda$ such that $\hat H_\Lambda^G \succ0$. In case of a positive answer, then at least one of the optimal control inputs yields an exponentially stable closed-loop system.
We ought to stress that the obtained solution is an arbitrary choice among the infinitely many optimal solutions, which depends on the initial pre-stabilization in a way which is possibly difficult to predict.

\section{Conclusions}
\label{sec:conclusions}

In this paper, we studied economic linear quadratic optimal control in case the optimal solution is not unique. We connected existing stability results for the linear and nonlinear case and we discussed how the quadratic strict pre-dissipativity assumption can be interpreted in terms of solutions of two optimal control problems formulated in forward and backward time. Furthermore, we provided  computationally viable procedures to verify if quadratic strict pre-dissipativity holds. Finally, we discussed the finite-horizon case and the case in which the non-unique control input solution results in non-unique state trajectories, such that some optimal solutions might be stabilizing while others not.

The interpretation of quadratic strict pre-dissipativity in terms of solutions of optimal control problems establishes a strong connection with the seminal paper~\cite{Willems1971} and suggests connections with some more recent contributions in economic MPC, e.g.,~\cite{Houska2015}. The extension of our interpretation of strict dissipativity to the full nonlinear case has been investigated in~\cite{Zanon2026a}.

\bibliographystyle{IEEEtran}
\bibliography{bibliography}

\end{document}